\documentclass[12pt]{article}
\usepackage{amssymb}
\usepackage{epsfig}

\begin{document}

\def\a{\alpha}   \def\b{\beta}    \def\c{\chi}       \def\d{\delta}
\def\D{\Delta}   \def\e{\epsilon} \def\f{\phi}       \def\F{\Phi}
\def\g{\gamma}   \def\G{\Gamma}   \def\k{\kappa}     \def\l{\lambda}
\def\L{\Lambda}  \def\m{\mu}      \def\n{\nu}        \def\r{\varrho}
\def\o{\omega}   \def\O{\Omega}   \def\p{\psi}       \def\P{\Psi}
\def\s{\sigma}   \def\S{\Sigma}   \def\vt{\vartheta} \def\t{\tau}
\def\vf{\varphi} \def\x{\xi}      \def\z{\zeta}      \def\th{\theta}
\def\Th{\Theta}
\def\ve{\varepsilon}

\def\CAG{{\cal A/\cal G}}
\def\CA{{\cal A}} \def\CB{{\cal B}}  \def\CC{{\cal C}}
\def\CD{{\cal D}} \def\CF{{\cal F}}  \def\CG{{\cal G}}
\def\CH{{\cal H}} \def\CI{{\cal I}}  \def\CL{{\cal L}}
\def\CO{{\cal O}} \def\CP{{\cal P}}  \def\CS{{\cal S}}
\def\CR{{\cal R}} \def\CT{{\cal T}}  \def\CU{{\cal U}}
\def\CW{{\cal W}} \def\CM{{\cal M}}  \def\CX{{\cal X}}
\newcommand{\be}{\begin{equation}}
\newcommand{\bea}{\begin{eqnarray}}
\newcommand{\ee}{\end{equation}}
\newcommand{\eea}{\end{eqnarray}}

\begin{titlepage}

\hfill\parbox{4cm} {KIAS-P01031\\ hep-th/0202191 \\ February 2002}
\vspace{15mm}
\baselineskip 8mm
\begin{center}
{\LARGE \bf Dynamical Aspects on \\ Duality between SYM and NCOS
\\from D2-F1 Bound State}
\end{center}

\vspace{10mm} \baselineskip 6mm

\begin{center}
Seungjoon Hyun$^a$\footnote{\tt hyun@phya.yonsei.ac.kr} and
Hyeonjoon Shin$^b$\footnote{\tt hshin@kias.re.kr}
\\[5mm]
{\it $^a$Institute of Physics and Applied Physics, Yonsei
University, Seoul 120-749, Korea \\ $^b$School of Physics, Korea
Institute for Advanced Study, Seoul 130-012, Korea \\
and \\
BK 21 Physics Research Division and Institute of Basic Science \\
Sungkyunkwan University, Suwon 440-746, Korea}
\end{center}
\thispagestyle{empty}

\vfill
\begin{center}
{\bf Abstract}
\end{center}
\noindent

 It has been shown that (2+1)-dimensional ${\cal N}=8$ super
 Yang-Mills (SYM) theory with electric flux is
related to (2+1)-dimensional noncommutative open string (NCOS)
theory by `2-11' flip. This implies that the instanton process in
SYM theory, which corresponds to D0-brane exchange(M-momentum
transfer) between D2-branes, is dual to the KK momentum exchange
in NCOS theory, which is perturbative process in nature. In order
to confirm this, we obtain the effective action of probe M2-brane
on the background of tilted M2-branes, which would correspond to
the one-loop effective action of SYM theory with non-perturbative
instanton corrections. Then we consider the dual process in NCOS
theory, which is the scattering amplitude of the wound graviton
off the D2-F1 bound state involving KK-momentum transfer in
$x^2$-direction. Both of them give the same interaction terms.
Remarkably they also have the same behavior on the nontrivial
velocity dependence. All these strongly support the duality
between those two theories with completely different nature.

\vspace{1cm}
\end{titlepage}


\baselineskip 6.5mm
\renewcommand{\thefootnote}{\arabic{footnote}}
\setcounter{footnote}{0}

\section{Introduction}
The ($p$+1)-dimensional matrix theory \cite{matrix,matrix-r} describes the
system of $N$ D$p$-branes wrapped on the $x^1, \cdots x^p$
direction in the scaling limit\footnote{$\alpha, \beta =
0,1,\cdots p$ denote longitudinal directions of the brane. Among
them the electric directions on the brane are denoted by $\mu, \nu
=0,1$ and the remaining directions of the brane are denoted by
$i,j =2,3 \cdots, p$. $I, J =p+1, \cdots , 9$ denote the
directions transverse to the brane. $M, N= 0,1, \cdots 9$ denote
ten dimensional coordinates, collectively.} \bea \alpha'\sim
\e^{1\over 2}~,\ \ \ g_s \sim \e^{3-p\over 4}~,\ \ \ g_{IJ}\sim
\epsilon~, \label{mscale} \eea while the metric components in the
longitudinal direction are kept fixed in the limit
$\epsilon\rightarrow 0$. In this scaling limit,
 all the closed string modes and
massive open string modes are decoupled and thus it becomes the
theory of massless open string modes only, which is $U(N)$ super
Yang-Mills (SYM) theory on the D-branes world-volume. One may note
that $U(1)$ multiplet in $U(N)=U(1)\times SU(N)/Z_N$ degrees of
freedom describe the overall motion of the system and are
decoupled from the rest degrees of freedom.

Our main concern in this paper is the world-volume theory of
D$p$-F1 bound states, in which fundamental strings are dissolved
into D$p$-branes and turn into electric flux on D$p$-branes. They
are half BPS states just like those of pure D$p$-branes. In this
case one may again take the same limit as (\ref{mscale}) which
gives ordinary $U(N)$ SYM theory with electric flux. The resultant
$SU(N)/Z_N$ theory with an $Z_N$ flux has a mass gap while the
remaining $U(1)$ theory is free\cite{bound}.

One can take different limit of the system, which is, so-called,
the NCOS limit\cite{ncos}. This scaling limit is achieved by
considering near critical electric field on D$p$-brane: \be 2 \pi
\a'\e^{01}F_{01}=1-{\e\over 2}~, \ee where critical electric field
corresponds to $\e=0$. The scaling of the background metric for
closed string is given by\cite{GMSS} \be g_{\m\n}= \eta_{\m\n}~, \
\ \ g_{ij} = \e  \delta_{ij}~, \ \ \ g_{IJ}=\e \delta_{IJ}~.
\label{met1} \ee The effective open string tension is $ {1\over 4
\pi \a'_e} ={\e \over 4\pi \a'}$. Therefore, while $\a'\equiv
l_s^2$ sets the scale of closed string modes, $\a'_e\equiv
l_e^2$ can be considered as the scale of open string modes
stretched in the electric direction.

In the presence of background electric field, or NS $B$-field on
the D-brane world-volume, the effective metric seen by the open
strings on the D-brane worldvolume is different from the metric
seen by bulk closed string modes. The effective open string
metric, noncommutativity parameter and effective open string
coupling $G_o^2$ can be determined as \cite{FT,SW} \be
G_{\m\n}=\e\eta_{\m\n}~, \ \ \  G_{ij}=\e \d_{ij}~, \ \ \
\Th^{\m\n} =2\pi\a'_e \e^{\m\n}~, \ \ \ G_o^2=g_s \e^{1\over 2}~.
\ee The NCOS limit \cite{ncos}, under which the bulk closed
string modes are decoupled, is given by $\e  \rightarrow 0$ while
taking $\a'_e$, $G_o$ fixed, and therefore is summarized as \bea
g_{\m\n}   \sim  \CO(1)~,\ \ \ g_{ij} \sim \e~,\ \ \ g_{IJ} \sim
\e~,\ \ \ g_s \sim \e^{-{1\over 2}}~,\ \ \ \a' \sim  \e~.
\label{nscale} \eea In this limit, the effective degrees of
freedom are those of open strings on noncommutative spacetime in
which
\[
[x_0, x_1]= \theta~.
\]

In general, these two limits are connected by
U-duality\cite{hyun}. From the case of D1-F1 bound states with the
scaling limits (\ref{mscale}) and (\ref{nscale}), one can easily
see that the (1+1)-dimensional SYM theory on $N$ D-strings with
$M$ units flux of electric field is S-dual to (1+1)-dimensional
NCOS theory of $M$ D-strings with $N$ units flux of electric
field\cite{GMSS,kle}. The scaling limits (\ref{mscale}) and
(\ref{nscale}) imply that the (2+1)-dimensional SYM theory  with
electric flux from D2-F1 bound states wrapped on two-torus is
related to (2+1)-dimensional NCOS theory by the, so-called,
`2-11' flip, i.e. circle compactifications along different
directions\cite{GMSS}.Further duality chain on each side relates
matrix theory and NCOS theory on higher dimensional
tori\cite{hyun}.

These dual relations between string theories on noncommutative
spacetime and ordinary gauge theories, though guaranteed from
dualities of `mother' M/string theory, are quite surprising.
Immediate question related on these dualities of different type of
theories is how to map various kinds of excitations in one theory
to those in its dual theory. In this paper, we are especially
interested in the dual of stringy degrees of freedom in NCOS
theory. In general, massive degrees of freedom of open strings are
non-BPS, and thus disappear into multi-particle states of massless
spectrum as the coupling becomes strong. In the cases of
(1+1)-dimensional and (2+1)-dimensional theories SYM theories are
S-dual to NCOS theories and thus the above arguments may apply.
However, in the (2+1)-dimensional theory on torus, we have
additional BPS states among string spectrum. They are states with
KK momentum along compactified $x^2$-direction. Under `2-11' flip,
they map to D0-branes, and thus become magnetic flux in the dual
SYM theory. The exchange of D0-branes between two D2-branes can be
interpreted as the instanton process in (2+1)-dimensional SYM
theory. In particular, it has been shown that $v^4$-terms and
their superpartners in the effective action of (2+1)-dimensional
${\cal N}=8$ SYM theory are completely determined by one-loop
contribution and instanton corrections\cite{pol029,pab119,hyu183}.
In this paper we focus on the same process in the SYM theory side
and consider the dual process in the NCOS theory to confirm the
dual relation between instanton process in SYM theory and KK
momentum exchange in NCOS theory. In section 2, we obtain the
effective action of probe M2-brane on the background of tilted
M2-branes, which correspond to the one-loop effective action of
SYM theory including non-perturbative instanton corrections. In
section 3 we compute the scattering amplitude of the wound
graviton off the D2-F1 bound state involving KK-momentum transfer
in $x^2$-direction. In section 4, we draw our conclusions. As for
a reference, the relation between these two theories are
summarized in table 1. 

\begin{center}
\begin{tabular}{|c|c|c|} \hline
 Theory & NCOS & SYM \\
\hline
(D2, F1)&$(M,N)$ &$(N,M)$\\
\hline
 $g_s$  & $  \e^{-{1\over 2}}G_o^2 $ & $
\e^{1\over 4} G_o^{-1}({r_2\over l_e})^{3\over 2} $ \\
  \hline
$\alpha'$ & $  \e\a'_e $&
$ \e^{1\over 2}{G_o^2 l_e^3\over r_2}$ \\
\hline gauge coupling $g_{YM}^2$ & ${G_o^2 \over l_e}$ &${ r_2^2
\over G_o^2 l_e^3}$
\\
 \hline
$x^2$ radius& $ \e^{1\over 2}r_2 $& $G_o^2l_e$ \\
 \hline
$x^{11}$ radius&$G_o^2l_e$& $\e^{1\over 2}r_2 $ \\
  \hline
&$k_2$& magnetic flux \\
 \hline
& magnetic flux &$k_2$ \\
\hline
$l_p^3$ &$\e G_o^2l_e^3 $ &$\e G_o^2l_e^3 $ \\
 \hline
\end{tabular}
\\
[.3cm]
 Table 1. 2+1 dimensional theories ($\alpha_e^\prime =l_e^2$ )
\end{center}

\section{Effective action of probe M2-brane}
In this section we consider D2-F1 bound state in the SYM theory
limit. In order to obtain the effective action of
(2+1)-dimensional ${\cal N}= 8$ SYM theory with electric flux, we
use the AdS/CFT correspondence\cite{adscft,ads-r}, which is
well-established in the matrix theory limit\cite{matrix-r,hyun98}.
Note that in this limit, the scale of transverse direction and the
scale of eleventh direction behave in the same way,
\[
r\sim \epsilon^{1\over 2}~, \ \ \ \ R\sim  \epsilon^{1\over 2}~,
\]
and hence we should take into account the contribution from
eleventh direction. D2-F1 bound state becomes tilted M2-branes
bound state in eleven-dimensional lift. Therefore we consider
probe M2-brane dynamics in the background of source tilted
M2-branes in the limit and obtain the effective action of probe
M2-brane which corresponds to the one loop effective action of
(2+1)-dimensional SYM theory with full non-perturbative instanton
corrections.

\subsection{Probe dynamics of M2-brane in the $AdS_4$ background}
In this subsection, we review the supergravity dual description of
(2+1)-dimensional ${\cal N}=8$ SYM theory, without electric flux.
In the limit (\ref{mscale}), the world-volume theory of $N$
D2-branes reduces to (2+1)-dimensional ${\cal N}=8$ SYM theory.
$v^4$ terms and their superpartners are completely determined by
one-loop and non-perturbative instanton corrections\cite{pol029}.
This is due to the non-renormalization property of the theory with
16 supercharges\cite{pab119}. On the other hand, the supergravity
dual description is given by the probe dynamics of M2-brane in the
background of periodically identified $AdS_4$ spacetime, produced
by the background D2-branes in the SYM decoupling
limit\cite{hyu183,che120}:
\begin{equation}
\label{ads4} ds^2_{11} = h_0^{-2/3} (-dt^2+dx_1^2+dx_2^2)
       +h_0^{1/3} (dx_3^2 + \cdots + dx_9^2+dx_{11}^2) ,
\end{equation}
where the eleven-dimensional harmonic function $h_0$ is given by
\begin{eqnarray}
\label{h0} h_0= \sum_{n = -\infty}^{\infty}
 \frac{2^5\pi^2l_p^6 N}{(r^2+ (x_{11} + 2 \pi R n)^2 )^3}~,
\end{eqnarray}
under $x^{11}$-compactification, $x_{11}\sim x_{11}+ 2\pi R$, and
 $r^2= x_3^2+\cdots  + x_9^2$.
Here $N$ denotes the number of background M2-branes.

We consider the probe dynamics of M2-brane wrapping on $x^1$,
$x^2$ directions and moving with a constant velocity
$v^I=\partial_0 x^I$ and $v^{11}=\partial_0 x^{11}$ in the
transverse directions. The bosonic part of the action for the
probe M2-brane is given by
\begin{equation}
\label{m2action} S_2 = -T_2 \int d^3\xi \; \sqrt{ - \det
h_{\alpha\beta} }
                   + i \mu_2 \int H  ~,
\end{equation}
where $h_{\alpha\beta}$ is the induced metric on the world-volume of
the probe M2-brane and is given by
\begin{equation}
h_{\alpha\beta} =  \partial_\alpha x^{\hat M}
\partial_\beta x^{\hat N} g_{{\hat M}{\hat N}}~,
\end{equation}
where ${\hat M}, {\hat N}$ denote full
eleven-dimensional spacetime coordinates, 0,1,$\cdots$ 9,11.

 From the configurations we choose, it is natural to use the static
gauge in which $x_\alpha=\xi_\alpha$. After plugging the metric
(\ref{ads4}) with the harmonic function $h_0$ given in (\ref{h0})
and expanding in powers of $v^2=v_I^2+ v_{11}^2$, the action
becomes
\begin{eqnarray}
S_2 = \int d^3 \xi \; \big({1\over 2} T_2 v^2 + {1\over
8}T_2h(v^2)^2 +{\cal O} ((v^2)^3)\big)~.
\end{eqnarray}

This effective action contains many informations on the dual SYM
theory. The vanishing effective action for $v^2=0$ tells that the
corresponding configuration is  BPS. In the dual SYM theory, 16
supercharge guarantees the non-renormalization of kinetic terms,
$v^2$, of bosonic fields, which is consistent with the above
action. Furthermore it has been shown that $v^4$-terms and their
superpartners are completely determined by one-loop and
non-perturbative instanton corrections. This can be shown to agree
with the above action as well, by using Poisson resummation
formula:
\begin{equation}
\sum^{\infty}_{n=-\infty} f(n) =
   \sum^{\infty}_{m=-\infty} \int^\infty_{-\infty} d \phi \;
   f( \phi ) \; e^{2 \pi i m \phi} ~,
\end{equation}
on the harmonic function $h_0$ which becomes
\[
  \sum_{n = -\infty}^{\infty}
  \frac{1}{(r^2+( x_{11} + 2 \pi R  n)^2 ) ^3}
 \  = \   \frac{1}{16 R} \ \Big[ \  \frac{3}{ r^5}
  + \frac{1}{ r^3}
      \sum^{\infty}_{m=1} \frac{m^2}{R^2} e^{-mr/R} 2
        \cos (mx_{11}/R)
\]
\begin{equation}
      + \frac{3}{r^4}
      \sum^{\infty}_{m=1} \frac{m}{R} e^{-mr/R} 2
      \cos (mx_{11}/R)
    + \frac{3}{r^5}
      \sum^{\infty}_{m=1} e^{-mr/R} 2 \cos (mx_{11}/R)~ \Big] .
     \label{resum}
\end{equation}
Note that the first term and the remaining terms with infinite sum
over the index $m$ correspond to the one-loop correction and the
instanton corrections, respectively, in the effective action of
(2+1)-dimensional SYM theory.

\subsection{The background geometry for D2-F1 bound state}
The bound state of D2-branes and fundamental strings is nothing
but the tilted M2 branes in eleven dimensions. The
eleven-dimensional geometry due to these tilted M2 branes are
given by the corresponding $x^2-x^{11}$ rotation of the above
metric (\ref{ads4})\cite{Dp-F1}:
\begin{equation}
\label{11dmetric} ds^2_{11} = h^{-2/3} (-dt^2+dx_1^2+dx_2^2)
       +h^{1/3} (dx_3^2 + \cdots + dx_9^2+(dx_{11}-\epsilon^{1/2}dx_2)^2) ,
\end{equation}
where the harmonic function $h$ becomes
\[ h= 1+ {2^5\pi^2l_p^6N
\over (r^2+ (\cos\theta x_2+ \sin\theta x_{11})^2)^3}~.\]

Under $x^{11}$-compactification, $x_{11}\sim x_{11}+ 2\pi R$, the
harmonic function $h$ can be written as
\begin{eqnarray}
h= 1+ \sum_{n = -\infty}^{\infty}
 \frac{2^5\pi^2l_p^6 N}{(r^2+\sin^2\theta( x_{11} + 2 \pi R n-
 \cot\theta x_2 )^2 ) ^3}~,
\end{eqnarray}
and also can be resummed using Poisson resummation formula as
follows:
\begin{eqnarray}
h-1 &=&   \frac{2\pi^2l_p^6 N}{ R\sin\theta} \ \Big[ \ \frac{3}{
r^5}
  + \frac{1}{ r^3} \sum^{\infty}_{m=1}
  (\frac{m}{R\sin\theta})^2 e^{-mr/R\sin\theta} 2
  \cos {m (x_{11}-x_2\cot\theta )\over R} \nonumber \\
& &+ \frac{3}{r^4}
  \sum^{\infty}_{m=1} \frac{m}{R\sin\theta} e^{-mr/R\sin\theta} 2
  \cos {m (x_{11}-x_2\cot\theta )\over R} \nonumber \\
& &+ {3 \over r^5}
  \sum^{\infty}_{m=1} e^{-mr/R\sin\theta}
  2 \cos {m (x_{11}-x_2\cot\theta )\over R}~ \Big]
\nonumber \\
&=& \frac{2\pi^2l_p^6 N}{R\sin\theta}
    \left[ \; \frac{3}{r^5} \right.
\\
&&\left.  +
       \sum^{\infty}_{m=1} \left( \frac{2}{\pi} \right)^\frac{1}{2}
              \frac{m^2 m^{1/2}}{(R\sin\theta)^5} \left(
         \frac{R\sin\theta}{r} \right)^{5/2}
                           K_{5/2} ( {mr \over R\sin\theta} )
        2 \cos{m (x_{11}-x_2\cot\theta )\over R} \;
    \right] \nonumber
\label{harmonic1}
\end{eqnarray}

In the $R\rightarrow 0$ limit, the geometry becomes the one
generated by the bound state of $N$ D2-branes and $M$ fundamental
strings:
\begin{equation}
ds^2 = \tilde{f}^{1/2} f^{-1} (-dx_0^2+dx_1^2)
      +\tilde{f}^{-1/2} dx_2^2 + \tilde{f}^{1/2}
      (dx_3^2 + \cdots + dx_9^2)~,
\label{10dmetric}
\end{equation}
where
\begin{equation}
f=1+\frac{r_0^5}{r^5}~,
\end{equation}
Here the rotation angle $\theta$ becomes
\[
\cos \theta = {g_s M {\sqrt{\alpha'} \over R_2} \over
 (g_s^2M^2{\alpha'\over R_2^2} + N^2)^{1/2}}
={MR \over (M^2 R^2 + N^2 R_2^2)^{1/2}}~,
\]
and the constant $r_0$ is given by
\[r_0^5 = 6\pi^2 g_s  \alpha'^{5/2}(g_s^2M^2{\alpha'\over
R_2^2} + N^2)^{1/2} = 6\pi^2 \alpha'{}^2{R\over R_2}(M^2 R^2+
N^2R_2^2)^{1/2}~,\] with $x^2$-compactification radius $R_2$.


\subsection{Probe M2-brane dynamics in the tilted M2-brane background}
\begin{figure}[!h]
\centerline{\hbox{\psfig{figure=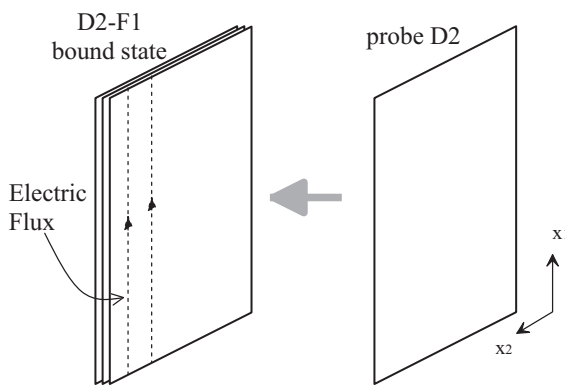,height=50mm}}}
\caption{{\footnotesize D2 scattering in SYM theory}} \label{d2f1}
\end{figure}

In this subsection we would like to get the effective action of
probe M2-brane in the background of tilted M2-branes, which is the
eleven-dimensional lift of probe D2-brane dynamics in the
background of source (D2-F1) bound state as shown in the Fig.
\ref{d2f1}.

The probe M2-brane action is given by (\ref{m2action}), now with
new geometry (\ref{11dmetric}). In order to recover the results of
SYM theory, we need to take the SYM limit (\ref{mscale}). In this
limit, $u= {r\over \alpha'}$ and $\phi_8={ x_{11}\over \alpha'}$
fixed, and we have\footnote{ Note that we use static gauge, $x_2=
\xi_2$.}
\begin{eqnarray}
l_p^3 h &=&  2\pi^2 N g_{YM}^4 \ \Big[ \  \frac{3}{ u^5}
  + \frac{1}{ u^3} \sum^{\infty}_{m=1}
  (\frac{m}{g_{YM}^2})^2 e^{-mu/ g_{YM}^2} 2
  \cos ({m \over  g_{YM}^2}(\phi_8-
{g_{YM}^2 \over R_2}{M \over N} \xi_2) )
 \nonumber \\
& & +\frac{3}{u^4}
  \sum^{\infty}_{m=1} \frac{m}{g_{YM}^2} e^{-mu/ g_{YM}^2} 2
  \cos ({m \over  g_{YM}^2}(\phi_8-
{g_{YM}^2 \over R_2}{M \over N} \xi_2) )
 \nonumber \\
& & + {3 \over u^5}
  \sum^{\infty}_{m=1} e^{-mu/ g_{YM}^2}
  2 \cos ({m \over  g_{YM}^2}(\phi_8-
{g_{YM}^2 \over R_2}{M \over N} \xi_2) )~ \Big] \label{harm2}
\end{eqnarray}
which is finite under the limit $\epsilon\rightarrow 0$.

Here $\phi_8$ is interpreted as the dual scalar of
(2+1)-dimensional gauge field $A_\mu$,
\[
\partial_\mu \phi_8 = \varepsilon_{\mu\nu\rho}F^{\nu\rho}~.
\]
 In the D2-F1 bound state, fundamental strings are dissolved
and turned into electric flux, i.e. $F_{01}^{(B)}\neq 0$. This
implies the nontrivial background value of the dual scalar
$\phi_8^{(B)}\propto \xi_2$ as
\[
\partial_2 \phi_8^{(B)}\neq 0~.
\]
This is gauge theory side interpretation why we have extra term in
the argument of cosine function in the eq. (\ref{harm2}), namely,
we can set
\begin{eqnarray}
\phi_8^{(B)}=- {g_{YM}^2 \over R_2}{M \over N} \xi_2
\label{back}
\end{eqnarray}.

The $M$ unit of electric flux in $N$ D2-branes can be expressed as
\begin{eqnarray}
M= N{2\pi R_2\over g_s} { (\a')^{1/2}\e^{01}F_{01}^{(B)}\over
\sqrt{1-(2\pi \a')^2 F_{(B)}^2}}~. \label{e-field}
\end{eqnarray}
Therefore, in the SYM limit, the background electric field is
given by
\[
F_{01}^{(B)}={g_{YM}^2 \over 2\pi R_2}{M\over N}~, \] which is
consistent with the above assignment (\ref{back}):

The effective action of probe M2-brane in the geometry
(\ref{11dmetric}) becomes
\begin{eqnarray}
S_2 &=& -T_2 \int d^3\xi \; {1\over h
}\big(\sqrt{(1+h\epsilon)(1-h\epsilon {\bar v}_I^2-h\epsilon {\bar v}_{11}^2)+
h^2\epsilon^2 {\bar v}_{11}^2 } - 1\big)\nonumber \\
 &=& -\frac{T_2\epsilon}{2} \int d^3\xi \;
   \bigg[
     (1-{\bar v}_I^2-{\bar v}_{11}^2) \nonumber \\
  &&\; \; \; \ \ \ \   - \frac{1}{4} h\epsilon
(1+{\bar v}_I^2+{\bar v}_{11}^2+2{\bar v}_{11})
     (1+{\bar v}_I^2+{\bar v}_{11}^2-2{\bar v}_{11})
   + {\cal O} ((h\epsilon)^2)
   \bigg]~,
 \label{pd}
\end{eqnarray}
where   $h$ is given by (\ref{harm2}) and ${\bar v}_I= {v_I \over
\epsilon^{1/2}}$ and ${\bar v}_{11}= {v_{11} \over \epsilon^{1/2}}$
are fixed under the
limit. One should note that in the present case the effective
action is nonvanishing even in the case for the vanishing
velocity, $v=0$, as it is not a supersymmetric configuration.

In order to compare with the results of NCOS theory in the next
section, we rewrite (\ref{harm2}) in terms of NCOS variables using
the relation shown in table 1 as follows:

\begin{eqnarray}
l_p^3 h
 &=&  6\pi^2 N {\epsilon^{5\over 2}G_o^6 l_e^9\over r_2 r^5} \ \Big[ \
\big(1+ \sum^{\infty}_{m=1}e^{-{mr \over
r_2 \epsilon^{1/2}}}  2\cos ({m x^{11}\over  r_2\epsilon^{1/2}}-
{m \xi_2 \over r_2})\big) \nonumber \\
  & & +
 \sum^{\infty}_{m=1}{m r \over r_2 \epsilon^{1/2}}
e^{-{mr \over
r_2 \epsilon^{1/2}}} 2\cos ({m x^{11}\over  r_2\epsilon^{1/2}}-
{m\xi_2 \over r_2}) \nonumber
\\
&& +\frac{1}{3}
\sum^{\infty}_{m=1}
  ({m r \over r_2 \epsilon^{1/2}} )^2 e^{-{mr \over
r_2 \epsilon^{1/2}}} 2
  \cos ({m x^{11}\over  r_2\epsilon^{1/2}}-{m \xi_2\over r_2})
~ \Big]
\label{harm3}
\end{eqnarray}

In the next section we would like to recover the above results
from the dual NCOS theory. In particular, we will obtain exactly
the same form as the terms linear in the harmonic function $h$ in
(\ref{pd}).

\section{Scattering of fundamental string off D2-F1 bound state
in NCOS theory} 
\begin{figure}[!h]
\centerline{\hbox{\psfig{figure=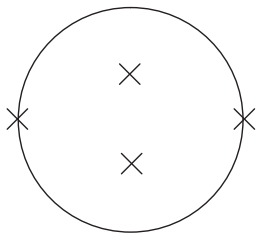,height=30mm}}}
\caption{{\footnotesize Process in the NCOS theory dual to
M-momentum transfer in the SYM theory}} \label{6pt}
\end{figure}

We now turn to the NCOS theory and consider the
process dual to that of D0-brane exchange between D2-branes. The
D0-brane exchange is interpreted as the momentum transfer in the
eleventh dimention, M-momentum transfer, which becomes the
momentum exchange in the $x^2$ direction in the NCOS theory side,
through the `2-11' flip or T$_2$ST$_2$ duality chain.  The probe
D2-brane  in the SYM side corresponds to a fundamental string in
the NCOS theory under the same duality chain.  The dual process of
the probe D2-brane scattered off the source D2-brane is then given
by the usual string amplitude for the scattering of fundamental
closed string off the D-brane. The diagram for the amplitude is
depicted in Fig.\ref{6pt}.  It has two closed string vertices and,
in addition to them, two open string vertex insertions, which are
necessary for describing the momentum transfer in the $x^2$
direction.

The process in the SYM side considered in the last section is the
low energy one and assumes that the branes have no fluctuating
modes on their worldvolume.  This leads us to take the vertices in
the string diagram for the dual process to be those for lowest
states, that is, massless states.  For the closed string, we take
graviton states polarized transversely to the D2-brane to consider
the process dual to that in the SYM side.  Associated with the
momentum transfer in the $x^2$ direction, the vector states
polarized along the D2-brane world-volume directions are taken as
open string vertices, since the dual process in the SYM side is
represented by the gauge field dynamics.  Note that the D2-brane
carrying electric flux spans $x^1$ and $x^2$ spatial directions,
which are compactified on circles of radius $R_1$ and $R_2$,
respectively. According to the duality chain, closed string dual
to the probe D2-brane in the SYM side has winding along the $x^1$
direction. In view of the NCOS limit, this is the right situation,
because only closed string winding along the direction where the
electric field is turned on (here $x^1$) can be involved in the
NCOS dynamics
\cite{kle085}.

For the evaluation of the disk diagram, Fig.\ref{6pt}, we use the
usual doubling trick \cite{has214} which expresses the world-sheet
field $X(z,\bar{z})$ in terms of its holomorphic part only as
follows:
\begin{equation}
X^M (z,\bar{z}) = X^M (z) + (D M^{-1})^M{}_N X^N (\bar{z})~,
\end{equation}
where $D$ is the diagonal matrix with elements $+1$ in the
directions parallel to the D2-brane and $-1$ in transverse
directions.  The matrix $M$ is due to the boundary conditions in
the presence of the electric field $E$ and its non-trivial part is
given by \footnote{A detailed discussion for the matrices $M$ and
$D$ may be found in \cite{hyu059}.}
\begin{equation}
(M^{-1})^\mu{}_\nu = \frac{1}{1-E^2}
   { \left(
    \begin{array}{cc}
        1+E^2 & 2 E \\
         2 E & 1+E^2
    \end{array}
    \right)^\mu}_\nu ~.
\end{equation}
For other directions, $M$ is identity.  From now on, we represent
$DM^{-1}$ as $R$ for notational convenience;
\[
R \equiv DM^{-1}~.
\]

The string scattering amplitude we will consider is then given by
\begin{equation}
{\cal A} \simeq \int d^2z d^2 z' \int dy dy' \langle
V(z,\bar{z}),V'(z',\bar{z}') V(y) V'(y') \rangle~.
\end{equation}
where the first two vertices are for wound
gravitons\footnote{Detailed description for the wound graviton can
be found in \cite{dan}.} and the last two vertices for
vector fields on the D2-brane. Each vertex is expressed as
follows:\footnote{The factor $\epsilon
\equiv 1-E^2$ multiplied to the vertex operator in the $-1$
picture for the wound graviton was absent in a previous literature
\cite{her017}.  This is due to the fact that we use the metric
$g^{MN}$ in the Green's function for the world-sheet fermions,
Eq.(\ref{green}), rather than $\eta^{MN}$.}
\begin{eqnarray}
V(z,\bar{z}) &=& G_o^2 \epsilon
    \xi_{IJ} :V^I_{-1} (p,z)::V^J_{-1}
     ( R^T \tilde{p},\bar{z}):~,
        \nonumber \\
V'(z',\bar{z}') &=& \frac{G_o^2}{\alpha'_e}
    \xi'_{IJ} :V^I_0 (p',z'): :V^J_0
    ( R^T \tilde{p}',\bar{z}'):~,
        \nonumber \\
V(y) &=& \frac{G_o}{\sqrt{\alpha'_e}}
        \zeta \cdot (1+M^{-1})_\alpha
        :V^\alpha_0 (k \cdot (1+R), y):~,
        \nonumber \\
V'(y') &=& \frac{G_o}{\sqrt{\alpha'_e}}
        \zeta' \cdot (1+M^{-1})_\alpha
        :V^\alpha_0 (k'\cdot(1+R),y'):~,
\end{eqnarray}
where $p$ and $p'$ ($\tilde{p}$ and $\tilde{p}'$) are momenta of
the wound gravitons contributed from the left (right) moving
sector of the closed string. The vertex operators in each picture
are given by
\begin{eqnarray}
V^M_{-1}(q,z) &=& e^{-\phi(z)}
        \psi^M (z) e^{i q \cdot X(z)}~,
        \nonumber \\
V^M_0 (q,z) &=&
        \left( \partial X^M (z)
         + i \frac{\alpha'}{2} q\cdot \psi (z)
          \psi^M(z)
        \right) e^{i q \cdot X (z)}~.
\end{eqnarray}

The evaluation of the amplitude ${\cal A}$ in a fully covariant
way is a formidable task and is not our aim.  Since we are
interested in the process dual to the M-momentum transfer between
D2-branes in the SYM side, we now take a particular situation.
Firstly, we take the backward scattering at least in the $x^2$
direction, since, in the SYM side, the D2-branes do not cross each
other in the eleventh direction while transferring M-momenta.
Secondly, all the momenta are assumed not to have KK momenta in
the $x^1$ direction, which is also the case in the SYM side.

Then, for the open string states, the momenta $k$ and $k'$ are
taken to be $k_\alpha = (m/R_2,0,m/R_2)$ and $k'_\alpha =
(-m'/R_2,0,m'/R_2)$, where $m$ and $m'$ are integers.  Since the
open string states are massless and thus satisfy $\zeta \cdot k =
\zeta' \cdot k' = 0$, and $k^2=k'^2=0$, the polarizations can be
chosen to have no components in the $x^0$ and $x^2$ directions.
The closed string states have winding in the $x^1$ direction.  Two
wound gravitons are set to have equal winding number.  In the
$x^2$ direction, we take the states to have KK momentum but no
winding.  The components of each momentum are then as follows:
from the left moving part of the closed string, $p_M = (p_0, R_1
w/ \alpha', n / R_2, p_\perp)$ $(p'_M = (-p'_0, -R_1 w /
\alpha', n' / R_2, -p'_\perp))$ where $p_\perp$ $(p'_\perp)$
represents the momentum components transverse to the D2-brane and
$n$, $n'$, and $w$ are integers ; from the right moving part,
$\tilde{p}_M = (p_0, -R_1 w/ \alpha', n / R_2, p_\perp)$
$(\tilde{p}'_M = (-p'_0, R_1 w / \alpha', n' / R_2, -p'_\perp))$.
Since the closed string states are wound gravitons which are taken
to be polarized transversely to the D2-brane, we have $\xi^T=\xi$,
${\rm Tr} \xi = {\rm Tr} \xi' =0$, $\xi \cdot p =\xi \cdot
\tilde{p}=0$, and $\xi' \cdot p' = \xi' \cdot \tilde{p}'=0$.

Under the situation taken as above, what we are interested in is
the amplitude when the transverse momentum difference between the
two wound gravitons is very small, $|p'_\perp-p_\perp| \simeq 0$.
It is obtained by looking at the pole terms in $(p'+p)^2$ and its
evaluation is performed in a standard way, and thus we will omit
the calculational details.\footnote{Similar calculation with just
transverse momentum $(p_\perp)$ transfer has been done in
\cite{kri054}, in the context of NCOS theory as well as
non-relativistic string theory  \cite{oogo}.}
However, we would like to note two points: we
should bear in mind the momentum conservation law,
\begin{equation}
p+R^T \cdot \tilde{p} + p' + R^T  \cdot \tilde{p}' + k \cdot (1+R)
+ k' \cdot (1+R) = 0~, \label{momc}
\end{equation}
and the phase factors coming from the space-time noncommutativity
do not appear. The latter point implies that the Green's functions
for the world-sheet fields that we need are the usual ones, that
is,
\begin{eqnarray}
\langle X^M (z) X^N (z') \rangle
 &=& - \frac{\alpha'}{2} g^{MN} \ln (z-z')~, \nonumber
\\
\langle \psi^M (z) \psi^N (z') \rangle
 &=& - \frac{g^{MN}}{z-z'}~.
\label{green}
\end{eqnarray}

After fixing the $SL(2,{\bf R})$ symmetry present in the amplitude
${\cal A}$ which we choose as $z'=i$ and $y'=\infty$ and doing
some manipulations, we obtain the expression for the amplitude as
\begin{equation}
{\cal A} \simeq \frac{G_o^4}{\alpha'_e}
    \frac{1}{\alpha' (p + p')^2} {\rm Tr}
    (\xi \cdot \xi')(\zeta \cdot \zeta' )
    \alpha' k \cdot (p'-p) \alpha' k' \cdot (p'-p)
     + {\cal O} (p'+p)~,
\label{amp0}
\end{equation}
where the indices of polarization tensors are contracted with the
usual Minkowskian metric $\eta^{MN}$ while the other contractions
are done by using $g^{MN}$.  The part of ${\cal O} (p'+p)$ leads
to the interactions, more or less, corresponding to the
spin-dependent interactions in the dual SYM theory\cite{har039}.
Such kind of interactions is beyond of our concern because as we
are studying the process dual to that in the SYM theory, which is
spin-independent.

Now, we plug the components of momenta specified above into ${\cal
A}$, Eq.(\ref{amp0}), and eliminate $p_0$ ($p'_0$) through the
mass shell condition $p^2=0$ ($p'^2=0$).  Then, in the NCOS limit,
we get
\begin{equation}
{\cal A} \simeq
 \frac{G_o^4}{\alpha'_e}
 \frac{{\rm Tr}(\xi \cdot \xi') (\zeta \cdot \zeta')
        }{\alpha'_e (\Delta n^2/R_2^2 + q^2)}
 \frac{R_1^2 m^2 w^2}{R_2^2}
 (1+v_\perp^2 + v_{2}^2 +2 v_{2})(1+v_\perp^2 + v_{2}^2 -2 v_{2})
  + {\cal O} (p'+p)~.
\label{amp}
\end{equation}
where $\Delta n = n'+n$ is the amount of KK momentum exchange in
the $x^2$ direction and $q=p'_\perp-p_\perp$ is the momentum
transfer between the closed string and the D2-brane in the
transverse directions to the D2-brane.  The $v$'s are the
`velocities' of the closed string which are defined as
\begin{equation}
v_\perp = \frac{\alpha'_e}{R_1 w} p_\perp~,~~~ v_2
=\frac{\alpha'_e}{R_1 R_2 w} n~,
\end{equation}
where ${R_1 w\over \alpha_e^\prime}$ plays the role of
non-relativistic mass \cite{oogo}.

Let us turn to the probe dynamics in the SYM theory and expand the
potential in term of $\epsilon h$.  The leading interaction term,
which is linear order in $h$, Eq. (\ref{pd}), should be compared
with the amplitude (\ref{amp}).  We now see that the velocity
factors agree exactly with those in the SYM theory, if we perform
replacements, $v_\perp \rightarrow {\bar v}_I$ and $v_2
\rightarrow {\bar v}_{11}$.  As for the interaction type, there
should be terms in the NCOS theory that have the same form as
those in the harmonic function $\epsilon h$ in (\ref{harm3}), as
was alluded in the last section. In order to confirm this, we
Fourier transform the $1/(\Delta n^2/R_2^2 + q^2)$ factor in
Eq.(\ref{amp});
\begin{eqnarray}
&&\lefteqn{ \frac{1}{R_2} \sum_{\Delta n} e^{i \Delta n x_2/R_2}
\int {d^7q \over (2\pi)^7} e^{i q \cdot x_\perp}
\frac{1}{\alpha'_e (\Delta n^2/R_2^2 + q^2)} } \nonumber
\\
&=&
 \frac{3}{16\pi^3\alpha'_e R_2} \frac{1}{r^5}
 \bigg[\Big(1+ \sum^{\infty}_{\Delta n=1}e^{-\Delta n r / R_2}
2\cos (\Delta n x_2 / R_2)\Big)+\sum^{\infty}_{\Delta n=1}{\Delta
n r\over R_2} e^{-\Delta n r / R_2} 2\cos (\Delta n x_{2}/ R_2)
                                    \nonumber \\
 & & + {1\over 3}\sum^{\infty}_{\Delta n=1}
  \Big({\Delta n r\over R_2}\Big)^2
e^{-\Delta n r / R_2} 2\cos (\Delta n x_{2}/ R_2)
 \bigg]
                                    \nonumber \\
 &=& \frac{1}{16\pi^3\alpha'_e R_2} \frac{1}{r^5}
  \bigg[ 3 + \sum^\infty_{\Delta n=1} \Big( \frac{2}{\pi}
  \Big)^{1/2} \Big( \frac{\Delta n r}{R_2} \Big)^{5/2}
  K_{5/2}\Big( \frac{\Delta n r}{R_2} \Big) 2\cos (\Delta n x_{2}/ R_2)
 \bigg]~,
\end{eqnarray}
where $r = \sqrt{x_\perp^2}$.  This is exactly the same, up to an
overall coefficients, as the harmonic function $h$ in
(\ref{harm3}). Note that, in the comparison, `2-11' flip has been
implied, in which the KK-momentum exchange, $\Delta n$, in NCOS
theory is traded to the instanton process, i.e. D0-brane exchange,
$m$. This completes the study of process in the NCOS theory dual
to that in the SYM theory.

\section{Conclusions}
The duality between (2+1)-dimensional ${\cal N}=8$ SYM theory with
electric flux and (2+1)-dimensional NCOS theory is inherited from
`2-11' flip of 'mother' M/string theory. Still yet, this duality
is quite surprising as those two theories have completely
different characteristics. One is the ordinary gauge theory whose
low energy excitations are gauge fields, while the other is theory
of strings living on noncommutative spacetime. The duality is more
or less strong-weak duality. Therefore the general stringy
excitations in NCOS theory decay into massless states or stable
massive BPS states in the strong coupling limit, and thus can not
be seen in the dual SYM theory.

However, we can still find some pieces of evidence of the duality
by considering the process involving non-trivial BPS spectrum. One
 such a process in SYM theory is the one corresponding to D2-D2
scattering process which involves instanton contributions, i.e.
D0-brane (or M-momentum, in eleven-dimensional sense) exchange.
The dual process in NCOS theory is the KK-momentum transfer in
$x^2$-direction, in the scattering amplitude of the closed
fundamental string off the D2-F1 bound state. We showed that both
of them give rise to the same results in the linear terms in the
harmonic function $h$, up to overall coefficients. They have the
same behavior in the $r$ dependence as well as in the non-trivial
velocity dependence. In order to obtain infinite sum of instanton
corrections in the dual NCOS theory, we should sum over all the
KK-momentum transfer in $x^2$-direction. Higher order terms in $h$
in the effective action (\ref{pd}) are expected to correspond to
higher loop corrections of the same scattering amplitude in NCOS
theory.

In (3+1)-dimensional case, NCOS theory is S-dual to noncommutative
Yang-Mills theory. Furthermore NCOS theory on $T^2$ is U-dual
($T^2ST^2$-dual) to ${\cal N}=4$ SYM theory with electric flux.
Interestingly enough, the gauge coupling in SYM theory is
independent of the open string coupling in NCOS theory, and thus
it is not clear what happen to the stringy degrees of freedom in
the regime of SYM theory. We hope to return this issue in the near
future \cite{soon}.

\section*{Acknowledgments}
We would like to thank Sangmin Lee for useful discussions. The
work of S.H. was supported in part by grant No. 2000-1-11200-001-3
from the Basic Research Program of the Korea Science and
Engineering Foundation.

\end{document}